\documentclass{mem}
\usepackage{natbib}\usepackage{txfonts}\usepackage{balance}
\usepackage{graphicx}
\usepackage[a4paper]{hyperref}
\idline{79}{1}
\begin{document}
\newcommand{\Mo}{{\rm M_\odot}}
\newcommand{\kms}{\>{\rm km}\,{\rm s}^{-1}}

\title{
Formation of Nuclear Disks and Supermassive Black Hole Binaries in Multi-Scale Hydrodynamical 
Galaxy Mergers 
}

\author{L. \,Mayer,\inst{1,2} S. \,Kazantzidis,\inst{3} \and A. \,Escala\inst{4}}

\offprints{L. Mayer}

\institute{Institute for Theoretical Physics, University of Z\"urich, Winterthurestrasse 
190, CH-8057 Z\"urich, Switzerland
\and
Institut f\"ur Astronomie, ETH Z\"urich, Wolfgang-Pauli-Strasse 16, CH-8093 Z\"urich, Switzerland
\and 
Center for Cosmology and Astro-Particle Physics, The Ohio State University, 191 West 
Woodruff Avenue, Columbus, OH 43210, USA
\and
Kavli Institute for Particle Astrophysics and Cosmology, Stanford University, P.O. Box 20450, MS 29, 
Stanford, CA 94309 USA
}

\authorrunning{Mayer et al.}

\titlerunning{Nuclear Disks and SMBH Binaries in Hydrodynamical Galaxy Mergers} 

\abstract{

We review the results of the first multi-scale, hydrodynamical simulations  
of mergers between galaxies with central supermassive black holes (SMBHs)
to investigate the formation of SMBH binaries in galactic nuclei. We demonstrate that strong gas inflows 
due to tidal torques produce nuclear disks at the centers of merger remnants whose properties depend sensitively on the 
details of gas thermodynamics. In numerical simulations with parsec-scale spatial
resolution in the gas component and an effective equation of state appropriate for a starburst
galaxy, we show that a SMBH binary forms very rapidly, less than a million 
years after the merger of the two galaxies, owing to the drag exerted by the 
surrounding gaseous nuclear disk. Binary formation is significantly suppressed in the presence 
of a strong heating source such as radiative feedback by the accreting SMBHs.
We also present preliminary results of numerical simulations with
ultra-high spatial resolution of $0.1$~pc in the gas component. 
These simulations resolve the internal structure of the resulting nuclear disk down to parsec scales and 
demonstrate the formation of a central massive object ($\sim 10^8 \Mo$) 
by efficient angular momentum transport due to the disk's extended spiral arms. 
This is the first time that a radial gas inflow is shown to extend to 
parsec scales as a result of the dynamics and hydrodynamics involved in 
a galaxy merger, and has important implications for the fueling of SMBHs. Due to the rapid formation 
of the central clump, the density of the nuclear disk decreases 
significantly in its outer region, reducing dramatically 
the effect of dynamical friction and leading to the stalling of the two 
SMBHs at a separation of $\sim 1$~pc. We discuss how the orbital decay of the 
black holes might continue in a more realistic model which incorporates star formation and the 
multi-phase nature of the ISM.

}

\maketitle

\section{Introduction}

In recent years, compelling dynamical evidence has indicated that supermassive
black holes (SMBHs) are ubiquitous in galactic nuclei (e.g., Ferrarese \& Ford 2005). 
According to the  standard modern theory of cosmological structure formation, the Cold Dark 
Matter (CDM) paradigm (e.g., Blumenthal et al. 1984), galaxies in the Universe grow 
through a complex process of continuous mergers and agglomeration of smaller systems. 
Thus, if more than one of the protogalactic 
fragments contained a SMBH, the formation of SMBH binaries during galaxy assembly will be 
almost inevitable (e.g., Begelman et al. 1980). 

In a purely stellar background, as the binary separation 
decays, the effectiveness of dynamical friction slowly declines, and the pair can become
tightly bound via three-body interactions, namely by capturing stars that pass close to the 
black holes and ejecting them at much higher velocities (e.g., Milosavljevi{\' c} \& Merritt 2001).
If the hardening continues to sufficiently small relative distances, gravitational wave emission
becomes the dominant source of orbital energy loss and the two SMBHs may coalesce in less than a Hubble time. 
However, the binary orbit may stop shrinking before gravitational radiation becomes relevant as there 
is a finite supply of stars on intersecting orbits (e.g., Berczik et al. 2005).

During the assembly of galaxies, especially at high $z$, their SMBHs likely 
evolve within gas-rich environments. Merging systems such as the Ultraluminous 
Infrared Galaxies (ULIRGs) NGC 6240 and Arp 220 harbor large 
concentrations of gas, in excess of $10^9 \Mo$, at their 
center, in the form of either a turbulent irregular structure or of 
a kinematically coherent, rotating disk (e.g., Downes \& Solomon 1998). 
Massive rotating nuclear disks of molecular gas are also ubiquitous in galaxies
that appear to have just undergone a major merger, such as Markarian 231 
(Davies et al. 2004). Gas dynamics may thus profoundly affect the pairing of SMBHs both during and 
after their host galaxies merge (e.g., Escala et al. 2004; Kazantzidis et al. 2005). 

Recent simulations of the orbital evolution of 
SMBHs within an equilibrium, rotationally-supported, gaseous disk have 
shown that dynamical friction against the gaseous background leads to the 
formation of a tightly bound SMBH binary with a final separation of 
$<1$~pc in about $10^7$~yr (Escala et al. 2005; Dotti et al. 2006; Dotti et al., these proceedings).
Here we review the results of high-resolution $N$-body + smoothed particle
hydrodynamics (SPH) simulations of mergers between galaxies with central SMBHs 
having enough dynamic range to follow the black holes from hundreds of kiloparsecs 
down to sub-parsec scales, bridging more than ten orders of magnitude in density. 

\section{Methods}

The aim of this study is to investigate the orbital evolution and pairing 
of SMBHs in multi-scale galaxy mergers in the hydrodynamical regime. 
A thorough description of our methods is presented in Kazantzidis et al. (2005) and Mayer et al. 
(2007, hereafter M07) and we summarize them here. First, we started with two identical spiral 
galaxies, comprising a disk of stars and gas with an exponential surface density distribution, 
a spherical, non-rotating Hernquist bulge, and a spherical and isotropic NFW 
dark matter halo. We adopted parameters from the Milky Way model A1 of Klypin 
et al. (2002) to initialize the galaxy models.

Specifically, the dark matter halo had a virial mass of $M_{\rm vir}=10^{12}\Mo$, a
concentration parameter of $c=12$, and a dimensionless spin parameter
of $\lambda=0.031$. The mass, thickness and resulting scale length of the disk
were $M_d=0.04 M_{\rm vir}$, $z_{0}=0.1 R_d$, and $R_d=3.5$~kpc,
respectively. The bulge mass and scale radius were $M_b=0.008 M_{\rm vir}$
and $a=0.2 R_d$, respectively. The halo was adiabatically contracted
to respond to the growth of the disk and bulge resulting
in a model with a central total density slope close to isothermal. 
The galaxy models were consistent with the stellar mass Tully-Fisher and 
size-mass relations. A softened particle of mass 
$2.6 \times 10^6 \Mo$ was placed at the center of the bulge 
to represent a SMBH. This choice satisfies the $M_{\rm BH}-\sigma$ 
relation (Kazantzidis et al. 2005). Lastly, the gas fraction, $f_{\rm g}$, 
was chosen to be $10\%$ of the total disk mass. We used a standard cooling 
function for a primordial mixture of atomic hydrogen and helium. 
We also shut off radiative cooling at temperatures below $2 \times 10^{4}$~K that is 
a factor of $\sim 2$ higher than the temperature at which atomic radiative 
cooling would drop sharply due to the adopted cooling function. 
With this choice we effectively take into account non-thermal, 
turbulent pressure to model the warm ISM of a real galaxy.

The galaxies were placed on parabolic orbits with pericentric distances 
that were 20\% of the halo virial radius ($r_{\rm peri} \sim 50$~kpc), 
typical of cosmological mergers (e.g., Khochfar \& Burkert 2006). The initial separation 
of the halo centers was twice their virial radii and their initial relative velocity was determined from the corresponding 
Keplerian orbit of two point masses. Each galaxy consisted of $10^5$ 
stellar disk particles, $10^5$ bulge particles, and $10^6$ dark matter particles. 
The gas component was represented by $10^5$ particles. We employed a gravitational softening of 
$\epsilon = 100$~pc for both the dark matter and baryonic particles of the galaxy, and $\epsilon=30$~pc
for the particle representing the SMBH.

During the interaction between the two galaxies, the relative separation of the black holes 
followed that of the galactic cores in which they were embedded. 
The merging galaxies approached each other several times as they sank into one another via dynamical friction. 
After $\sim 5$~Gyr, the dark matter halos had nearly merged and the two baryonic cores, separated by 
about $6$~kpc, continued to spiral down. As much as 60\% of the gas originally present in the 
galaxies was funneled to the inner few hundred parsecs of each core by 
tidal torques and shocks occurring in the repeated fly-bys between the two galaxies (e.g., Barnes \& Hernquist 1996).
Each SMBH was embedded in a rotating gaseous disk of mass $\sim 4 \times 10^8 \Mo$ and size of a 
few hundred parsecs which was produced by the gas inflow. 

Second, just before the last pericentric passage of the two merging galaxies, we adopted the technique of 
particle splitting to increase the gas mass resolution in the central region of the 
computational volume. By selecting a large enough volume for the 
fine grained region one can avoid dealing with spurious effects at the coarse/fine boundary, such as two-body heating
due to scattering by massive particles of the low-resolution region. 
We selected the volume of the fine-grained region to be large enough to quarantee that the 
dynamical timescales of the entire coarse-grained region were much longer than those corresponding to the refined region.
Specifically, we performed the splitting in a volume of $30$~kpc in radius at the point where the two galaxy 
cores were separated by only $6$~kpc. The new particles were randomly distributed according to the SPH smoothing 
kernel within a volume of size $\sim h_p^3$, where $h_p$ is the smoothing length of the parent particle. 
The velocities of the child particles were equal to those of their 
parent particle (ensuring momentum conservation) and so was their temperature, while each child particle was
assigned a mass equal to $1/N_{\rm split}$ the mass of the parent particle, where $N_{\rm split}$ is 
the number of child particles per parent particle. The mass resolution in the gas component was 
originally $2 \times 10^4 \Mo$ and became $\sim 3000 \Mo$ after splitting, for a total of $\sim 1.5$ 
million SPH particles. For the standard calculations, the softening of the gas particles was set to $2$~pc. 
We note that the local Jeans length was always resolved by $10$ or more SPH smoothing
kernels (e.g., Bate \& Burkert 1997) in the highest density regions of the refined simulations. The softening of the 
black holes was also reduced from $30$~pc to $2$~pc, while the softening of dark matter and stellar particles 
remained $100$~pc as they were not split in order to limit the computational burden.
Therefore, stellar and dark matter particles essentially provide a smooth background potential, while the 
computation focused on the gas component which dominates by mass the nuclear region.
All simulations were performed with GASOLINE, a multi-stepping, parallel Tree-SPH $N$-body 
code (Wadsley et al. 2004).

The radiation physics in the refined simulations was modeled via an 
``effective'' equation of state that accounts for the net balance of 
radiative heating and cooling. The value of the adiabatic index, $\gamma$, namely the 
ratio between the specific heats, is the parameter that controls the degree 
of dissipation in the gas. While the various cooling and heating 
mechanisms should be followed directly, this simple scheme allows us to 
investigate the effect of thermodynamics on the structure of the merger remnant 
and on the orbital decay of the black holes. Lastly, we tested that the transition 
between the two thermodynamic schemes used in the different parts of the simulation
did not introduce spurious fluctuations in the hydrodynamical variables (M07).

\section{Effects of thermodynamics on the orbital decay of SMBHs}

Calculations that include radiative transfer show that the 
thermodynamic state of a solar metallicity gas heated by a starburst 
can be well approximated by an ideal gas with adiabatic index $\gamma=1.3-1.4$ 
over a wide range of densities (Spaans \& Silk 2000). For the standard refined simulation 
discussed in the present work, we adopted $\gamma=7/5$. 

\begin{figure}[]
\resizebox{\hsize}{!}{\includegraphics[clip=true]{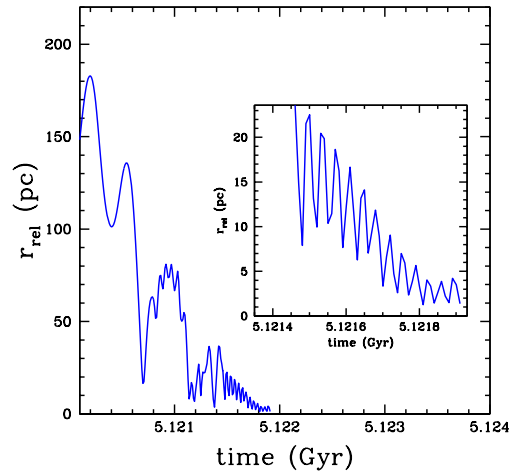}}
\caption{\footnotesize Relative separation of the two SMBHs as a function of time 
during the last stage of the standard, multi-scale merger simulation 
with $\gamma=7/5$. This value of $\gamma$ approximates well the
balance between radiative heating and cooling in a starburst
galaxy. The two peaks at scales of tens of parsecs at around $t=5.1213$~Gyr correspond to 
the end of the phase during which each black hole is still embedded in a
distinct gaseous core. The inset shows the details 
of the last part of the orbital evolution, which takes place inside the nuclear 
disk arising from the merger of the two galactic cores. A SMBH binary forms rapidly, 
less than a million years after the coalescence of the two galactic nuclei, 
owing to the drag exerted by the surrounding dense gaseous nuclear disk.
\label{fig1}}
\end{figure}
\begin{figure}[]
\resizebox{\hsize}{!}{\includegraphics[clip=true]{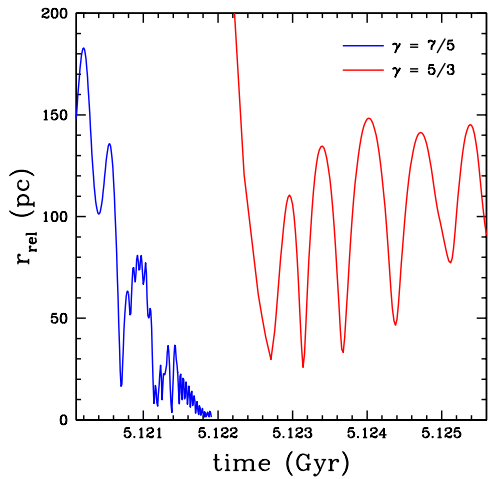}}
\caption{\footnotesize Relative separation of the two SMBHs as a function 
of time in two multi-scale, merger simulations with different prescriptions 
for the gas thermodynamics. The stiffer equation of state ($\gamma=5/3$) 
corresponds to a situation where radiative cooling is 
completely suppressed by a strong heating source (e.g., AGN feedback)
and causes the hardening process to significantly slow down.
The orbital decay and pairing of SMBHs depends sensitively on the 
details of gas thermodynamics.
}
\label{fig2}
\end{figure}

The gaseous cores finally merge at $t \sim 5.12$~Gyr, forming a single nuclear 
disk with a mass of $3\times 10^9 \Mo$ and a size of $\sim 75$~pc. The two
SMBHs are embedded in this nuclear disk. The disk is surrounded by several rings and by a more 
diffuse, rotationally-supported envelope extending out to more than
a kiloparsec. A background of dark matter and stars 
distributed in a spheroid is also present but the gas component is
dominant in mass within a few hundred pc from the center. From here on the orbital 
decay of the black holes is dominated by dynamical friction against this fairly dense 
gaseous disk. The black holes are on eccentric orbits and move with a speed of
$v_{\rm BH} \sim 200-300\kms$ relative to the disk's 
center of mass. The typical ambient sound speed is $v_s \sim 50\kms$.
The relative orbit of the SMBH pair decays from about $40$~pc to a few parsecs, our resolution limit, 
in less than a million years after the merger of the two galaxies (Figure~\ref{fig1}). 
At this point the two black 
holes are gravitationally bound to each other, as the gas mass enclosed within 
their separation is less than the mass of the binary. Dynamical friction against the 
stellar background would bring the two black holes this close only on a much longer 
timescale, $\sim 3 \times 10^7$~yr (Section 4). Such a short sinking timescale 
due to the gas is expected because of the high densities in the nuclear disk and 
because the decay occurs in the supersonic regime with $v_{\rm BH} > v_s$ (Ostriker 1999).
The subsequent hardening of the binary will depend on the details of gasdynamics and other 
processes at scales below the adopted resolution (Sections 6 \& 7).

It is interesting to investigate the effect of the adopted equation of state on the orbital decay 
of the black holes. In particular, we considered a smaller degree of dissipation in the gas  
and increased $\gamma$ to $5/3$. This value of $\gamma$ would correspond to a purely adiabatic 
gas, or equivalently to a situation where radiative cooling is completely suppressed.
The radiative feedback from an active galactic nucleus (AGN) is a good candidate for such
a strong heating source. In this case, we find that a turbulent, pressure supported cloud of a few hundred 
parsecs arises from the merger rather than a disk. The nuclear region is still gas dominated, but the 
gas mass is lower within $100$~pc relative to the $\gamma=7/5$ case. This is because of the adiabatic 
expansion of the gas following the final shock when the two cores merged.

Figure~\ref{fig2} demonstrates that the hardening process is significantly suppressed when a 
stiffer equation of state with $\gamma=5/3$ is adopted. In this case, the black holes do not form a binary 
and maintain a relative separation of $\sim 100-150$~pc well after the binary forms in the 
simulation with $\gamma=7/5$. The density of the gas in the nuclear region surrounding the SMBHs 
is a factor of $\sim 5$ lower compared to that in the $\gamma=7/5$ case, the sound speed is $v_s \sim 100\kms$, and the 
black hole velocity is $v_{\rm BH} \lesssim 100\kms$. The lower density and to a lesser extent the fact that 
the two black holes move subsonically ($v_{\rm BH} \lesssim v_s$) rather than supersonically greatly reduce 
the drag due to the gas distribution when $\gamma=5/3$ (Ostriker 1999). In Section~5, we briefly discuss 
how the structure and kinematics of the nuclear regions of merger remnants in the simulations 
with different values of $\gamma$ compare to those of observed systems.

Given the sensitivity of the black hole pairing process to the value of $\gamma$, a scenario in which 
the black holes rapidly form a binary owing to dynamical friction against the gas would require that 
AGN feedback has negligible thermodynamical effects on small scales. In fact,
this may be a more general requirement if the ubiquitous nuclear disk-like structures seen in many 
merger remnants are to be preserved. Interestingly, previous studies that included a prescription for AGN feedback
in similar galaxy merger simulations (e.g., Springel et al. 2005) find that feedback 
affects strongly the thermodynamics of the gas in the nuclear region only $>10^8$~yr after 
the galaxy merger is completed.

\section{Dynamical friction timescales}

It is important to examine if the black holes could still form a binary 
as a result of the interaction with the collisionless stellar background. 
Since the resolution of the collisionless components is likely inadequate 
to assess directly the effect of dynamical friction (Section 2), we opt to 
calculate the dynamical friction timescale in the collsionless background 
analytically (Colpi et al. 1999)
\begin{equation}
{\tau_{\rm DF}=1.2 {V_{\rm cir}r_{\rm cir}^2
\over GM_{\rm BH}\ln(M_{\rm sd}/M_{\rm BH})}\,\varepsilon^{0.4}} \ .
\label{dyn.friction}
\end{equation}
Here $V_{\rm cir}$ and $r_{\rm cir}$ are, respectively, the initial
orbital velocity and the radius of the circular orbit with the
same energy of the actual orbit of the black holes in the simulation, 
$\varepsilon$ is the circularity of the orbit, and $M_{\rm sd}$ is 
the sum of the dark matter and stellar mass within $r_{\rm cir}$. 

We calculate the decay time when the two black holes are separated by $100$~pc, 
that is at the periphery of the nuclear disk just after the galaxy merger. 
Drawing the numbers from the simulations, we have $r_{\rm cir} = 100$~pc, 
$V_{\rm circ}= 200\kms$,  $\varepsilon =0.5$,
$M_{\rm BH} = 2.6 \times 10^6 \Mo $ and $M_{\rm sd} = 5 \times 10^8 \Mo$. 
We find that the dynamical friction timescales in the collisionless background 
are equal to $5 \times 10^7$~yr and $3 \times 10^7$~yr in the $\gamma=5/3$ and $\gamma=7/5$ simulations, respectively
(the shorter timescale in the $\gamma=7/5$ case is due to the fact that the stars and halo contract 
adiabatically more in response to the higher gas mass concentration in this case, and hence $M_{\rm sd}$ is higher).
In comparison, the binary formation timescale in the simulation with $\gamma=7/5$ was only $5 \times 10^5$~yr 
(Figure~\ref{fig1}). 

We stress that eq.~(\ref{dyn.friction}) was derived for an isothermal sphere. The stellar and dark matter 
distribution are indeed only mildly triaxial within a few hundred parsecs from the center of the remnant
and the total density profile is fairly close to $\rho(r) \propto r^{-2}$, as expected from previous 
work (e.g., Kazantzidis et al. 2005). We also note that eq.~(\ref{dyn.friction}) actually yields 
a lower limit to the dynamical friction timescale since close to parsec scales, as the binary becomes 
hard, evacuation of the stellar background due to three-body encounters will take place and the efficiency 
of the sinking process will be greatly reduced. Whether orbital decay will
continue and eventually lead to coalescence of the two black holes is uncertain in this case.
Centrophilic orbits in triaxial systems could help in refilling the loss cone and decrease the binary's 
separation to the point where the emission of gravitational waves becomes efficient at extracting the 
last remaining angular momentum (Berczik et al. 2005). However, as we just mentioned, the structure of the stellar core
is only mildly triaxial. Further investigation with simulations having higher resolution in the collisionless
component is needed. The $\gamma=5/3$ run was stopped $5 \times 10^6$~yr after the merger of the gaseous cores 
is completed. Once again, the fact that there is no evidence that the black holes are sinking until the end is
likely due to insufficient mass and force resolution in the collisionless background that does
not allow to resolve dynamical friction properly.

We also compared our results with the {\it expected} dynamical friction timescale due to the gaseous 
background. In the simulation with $\gamma=7/5$, the gas is distributed in a disk rather than in
an isothermal sphere. Since the disk thickness is $> 10$ times the black hole gravitational 
softening and because of the fact that the density profile of the disk can be roughly 
approximated with a power law with an index close to 2 (except at the center where it becomes steeper) 
we are allowed to use eq.~(\ref{dyn.friction}) to obtain a rough estimate of the timescales.
As shown by Escala et al. (2004), analytical predictions with a fixed Coulomb logarithm (Ostriker 1999) can overestimate the drag in the 
supersonic regime by a factor of $\sim 1.5$. In the $\gamma=7/5$ simulation, the black holes move 
supersonically and the analytical formula should yield the correct prediction. In this case the drag is a factor of 
$\sim 2.3$ stronger than in the corresponding collisionless case (Escala et al. 2004).
This is fairly consistent with our results. Indeed, eq.~(\ref{dyn.friction}) with a reduction of 
a factor of $2.3$ gives $\sim 10^6$~yr if we set $M_{\rm gas}=M_{\rm sd}$, with $M_{\rm gas} \sim 20 M_{\rm stars}$.  
This timescale has to be compared with that measured directly in the simulation, $5 \times 10^5$~yr. As discussed above, 
the gas profile is actually steeper than $r^{-2}$ near the center. Thus, it is not surprising that the 
decay is faster. Despite the apparent agreement with the analytically estimated drag, we note 
that the orbital evolution of the two black holes might be affected by more than just the gravitational wake. 
Indeed, the nuclear disks show strong, highly dynamical non-axisymmetric structures such as spiral arms 
(see Section 7) which are highly efficient at removing angular momentum from the orbiting SMBHs.

The drag drops rapidly by an order of magnitude in the subsonic regime (Escala et al. 2004).
This coupled with the fact that $M_{\rm gas}$ is a factor of $\sim 5$ lower in 
the simulation with $\gamma=5/3$ compared to that with $\gamma=7/5$ would give 
a drag $50$ times smaller or $\tau_{\rm DF} \sim 5 \times 10^7$~yr, explaining why the 
orbital decay caused by the gas is so inefficient in this case. Thus, 
in the $\gamma=5/3$ simulation stars and gas contribute to the drag 
in a comparable way.

Adding star formation is unlikely to change the above conclusions in any significant way. 
The unrefined galaxy merger simulation yields a starburst timescale of $\sim 5 \times 10^7$~yr. 
During this time, which is much longer than the binary formation timescale in the run with $\gamma=7/5$, 
half of the gas in the nuclear disk will be turned into stars. Instead, due to the fact that the black hole 
sinking timescale is comparable to that of the star formation timescale in the $\gamma=5/3$ simulation, 
the overall orbital evolution will be dictated by the stars rather than by the gas.
There are, however, some caveats in the argument regarding the role of star formation in the $\gamma=7/5$
case. First, the starburst timescale is based on the unrefined merger simulations. Had we included
star formation in the refined simulations we would have probably found shorter timescales locally
since these simulations can resolve much higher densities and the star formation rate 
depends on the local gas density. Second, one might wonder how the inclusion of feedback from star
formation, which was neglected in the unrefined merger simulations, would affect gas properties 
and, consequently, the orbital decay of the black holes. We defer a detailed numerical study of these 
considerations to future work.

\section{Structure, kinematics, and gas inflow in the nuclear regions of merger remnants}

The nuclear disk produced in the $\gamma=7/5$ case is highly turbulent. The sources of turbulence
are the prominent shocks generated as the cores merge and the persistent non-axisymmetric structures 
sustained by the self-gravity of the disk after the merger is completed (e.g., Wada \& Norman 2002).
The perturbation due to the black hole binary is a negligible effect since its mass is about $10^3$ times 
smaller than the mass of the disk. The degree of turbulence,
of order $50-100\kms$ as measured by the radial velocity dispersion, is comparable to that of
observed circumnuclear disks (e.g., Downes \& Solomon 1998).
The disk is composed by a very dense, compact region of size about $25$~pc which contains
half of its mass (the mean density inside this region is $> 10^5$ atoms/cm$^3$). The
outer region instead, from $25$ to $75-80$~pc, has a density $10-100$ times lower, and is surrounded
by even lower-density rotating rings extending out to a few hundred parsecs. The disk scale
height also increases from inside out, ranging from $20$~pc to nearly $40$~pc.
The volume-weighted density within $100$~pc is in the range $10^3-10^4$ atoms/cm$^3$, comparable to 
that of observed nuclear disks (e.g., Downes \& Solomon 1998).
This suggests that the degree of dissipation implied by the equation 
of state with $\gamma=7/5$ is reasonable despite the simplicity of the thermodynamical 
scheme adopted.

The rotating, flattened cloud produced in the $\gamma=5/3$ is instead more turbulent and less
dense than observed circumnuclear disks in merger remnants. The mean velocity dispersion measured within
$100$~pc is about $300\kms$, higher than the mean rotational velocity within the same radius,
which is $\sim 250\kms$. This suggests that the  $\gamma=5/3$ simulation does not describe the
typical nuclear structure resulting from a dissipative merger.

\begin{figure}[]
\resizebox{\hsize}{!}{\includegraphics[clip=true]{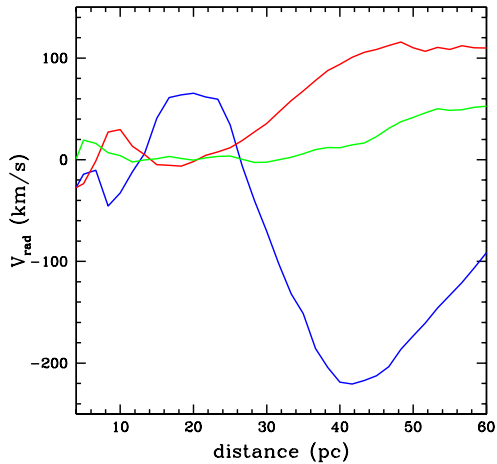}}
\caption{
\footnotesize Radial velocities inside the nuclear disk in the standard, multi-scale merger 
simulation with $\gamma=7/5$. The blue line corresponds to $t=5.1218$~Gyr, while
red and green lines show results after $10^5$~yr and $2 \times 10^5$~yr, 
respectively. Remarkable gas inflows and outflows are the result of streaming motions 
within the bar and spiral arms. These arise during the phases of strong, 
non-axisymmetric instabilities sustained by the disk self-gravity. 
At late times, the instabilities saturate due to self-regulation and the radial 
motions also decrease.
}
\label{fig3}
\end{figure}

The strong spiral pattern associated with the nuclear disk in the simulation with $\gamma=7/5$
produces remarkable radial velocities (Figure~\ref{fig3}). 
Since spiral modes transfer angular momentum inwards and mass outwards, strong inward radial velocities are 
expected. The amplitude of radial motions evolves with the amplitude of the spiral pattern; radial motions
decline as the spiral arms weaken over time. Just after the merger, when non-axisymmetry is strongest, 
radial motions reach amplitudes of $\sim 100\kms$ (Figure~\ref{fig3}).
This phase lasts only for a couple of orbital periods, while later the disk becomes smoother as spiral shocks increase the 
internal energy which in turn weakens the spiral pattern. Inward radial velocities of order $30-50\kms$ are seen
for the remaining few orbital times during which we are able to follow the system (Figure~\ref{fig3}).
Such velocities are comparable to those recently reported in high-resolution observations of nuclear disks 
of nearby Seyfert galaxies (Fathi et al. 2006). As the gas reaches down to a distance of few parsecs from the center, 
its radial velocity diminishes as we approach the resolution limit of the simulations ($\sim 2$~pc). 
Therefore, the fact that there is almost no net radial velocity within a few parsecs from the center 
(Figure~\ref{fig3}) is an artifact of the limited numerical resolution.

If we assume that speeds of $30-50\kms$ can be sustained down to scales of a few parsecs, more than
$10^8 \Mo$ of gas could reach parsec scales in about $10^5$~yr. This timescale is
much smaller than the duration of the starburst, and therefore such gas inflow should develop in a similar way 
even when star formation is taken into account. The inflow is also marginally faster than the decay timescale of the binary SMBH 
measured in the simulation ($\sim 5 \times 10^5$~yr). Presumably some of this gas 
could be intercepted by the two SMBHs as they are spiraling down (the relative velocities between the gas and the 
black holes are small since the SMBHs are always corotating with the nuclear disk)

\section{Merger simulations at sub-pc scales; nuclear fueling and orbital decay of SMBHs}

In the study of M07, the simulations were stopped when the two black holes had a separation of about $2$~pc,
comparable to the adopted gravitational softening length, which sets the nominal resolution limit. At this stage, 
the black holes had formed a loose binary on a fairly eccentric orbit. 
In order to explore the sinking of the SMBH binary to even smaller scales we performed a new simulation 
with a spatial resolution in the gas component of $0.1$~pc, that is $20$ times higher compared to 
M07. This resolution is comparable to the highest resolution achieved in simulations of 
nuclear disks starting from equilibrium initial conditions rather than from a large
scale galaxy merger (Escala et al. 2005; Dotti et al. 2007). On the other hand, the mass 
resolution in the new simulation was kept the same as in M07. The number of gas particles in the nuclear disk 
forming after the merger is sufficiently high ($\sim 10^6$) that even with $0.1$~pc resolution 
in the gas the Jeans mass is resolved by several SPH kernels in the disk, thus avoiding spurious numerical effects 
such as artificial fragmentation (Bate \& Burkert 1997).

\begin{figure}[]
\resizebox{\hsize}{!}{\includegraphics[clip=true]{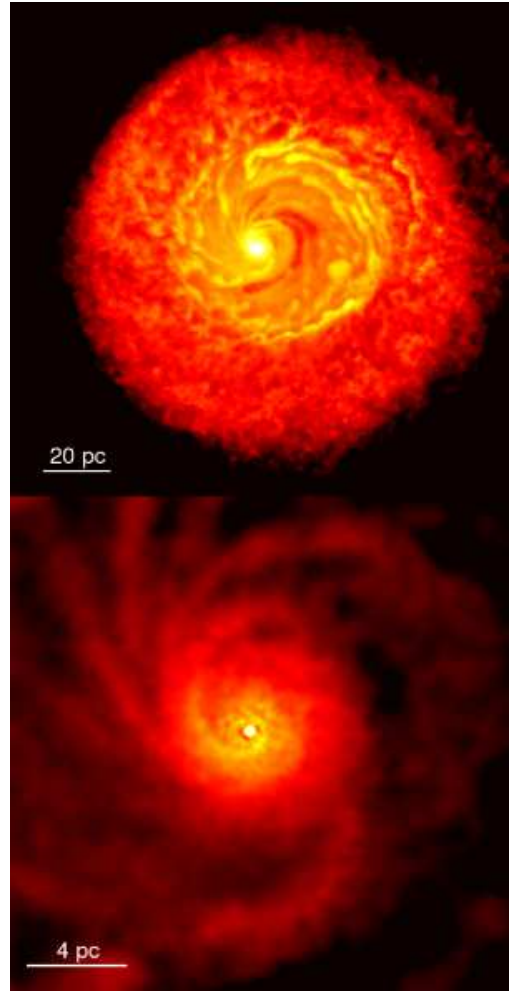}}
\caption{
\footnotesize
Color-coded projected density maps of the nuclear disk viewed face-on in 
the numerical simulation with $\gamma=7/5$ and $0.1$~pc spatial resolution
in the gas. Both panels display the nuclear disk $\sim 10^6$~yr after the galaxy merger is deemed complete.
The bottom panel presents the inner region of the disk. A conspicuous spiral pattern 
reaching to the central region and a central, massive clump produced by the 
strong gas inflow are evident.
}
\label{fig4}
\end{figure}

Due to the much higher spatial resolution, the nuclear disk now reveals a much richer structure with 
both large and small scale spiral patterns (Figure~\ref{fig4}). While the large scale spiral structure was
also reported in the simulations of M07, the inner few $10$~pc, that were quite featureless before, now 
reveal a high order spiral pattern extending down to sub-parsec scales. The spirals-in-spirals 
patterns are reminiscent of the bars-in-bars patterns that are suggested as possible candidates 
for bridging large and small scale inflows in non-interacting galaxies. 

\begin{figure}[]
\resizebox{\hsize}{!}{\includegraphics[clip=true]{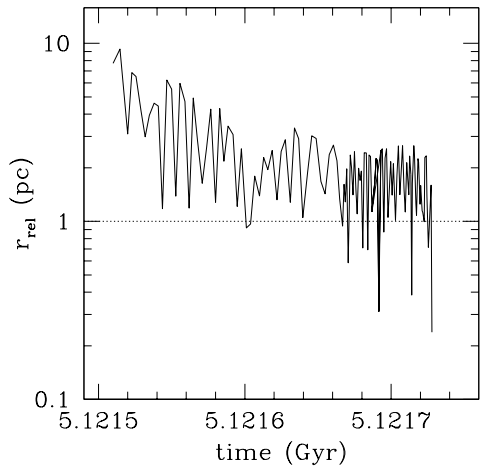}}
\caption{
\footnotesize Orbital evolution of the two SMBHs as a function of time in 
a multi-scale, merger simulation with a resolution of $0.1$~pc 
in the gas component. Results are presented for an equation of state 
with $\gamma=7/5$. The relative separation of the two SMBHs oscillates 
between $\sim 0.5$~pc and $\sim 2$~pc and never reaches the resolution 
limit of the simulation. As a result of very strong gas inflows, the 
density of the surrounding gas decreases considerably reducing the 
effect of dynamical friction on the SMBHs.
}
\label{fig5}
\end{figure}

Up to the point when the SMBHs reach a relative separation of $1-2$~pc the sinking rate is comparable to what 
was previously reported by M07. However, at smaller relative separations, the binary's orbital decay slows 
down and the orbit oscillates between a fraction of a parsec and $\sim 1$~pc (Figure~\ref{fig5}).
What causes this relative stalling? The answer lies in the evolution of the gas density and temperature 
profile in the nuclear disk. In a few $10^5$~yr, the strong gas inflow produces a very dense central clump with 
a mass of $\sim 10^8 \Mo$ and size of only $0.5$~pc. We note that this constitutes the first demonstration that 
a galaxy merger can produce remarkable concentrations of gas at sub-parsec scales subsequent to the formation of 
a nuclear disk from a larger scale gas inflow. While the density in the very
center goes up by an order of magnitude as the central clump forms, at scales above $1$~pc the
density decreases as angular momentum is transported outwards leading to the expansion of the disk. 
The disk exhibits a strong non-axisymmetric structure, with multi-armed spirals, extending down to the inner
few tens of parsecs (Figure~\ref{fig4}). The inner spiral pattern  is responsible for the efficient transfer of 
angular momentum in the inner disk region and is probably due to the SLING mechanism (Adams et al. 1989; 
Krumholz et al. 2007) which enables accretion on the disk orbital timescale rather than on the
viscous timescale. 


By the time the two black holes reach a separation of about $1$~pc the central clump has
already formed, sweeping most of the mass from the inner disk 
As a result of the central inflow, the density of the surrounding gas decreases by a factor of $\sim 5$. 
The density reduction weakens the effect of dynamical friction. We note that a similar
phenomenon is seen by Dotti et al. (these proceedings). In their simulations, the disk
profile becomes flatter rather than steeper, but it is still the case that the black holes
find themselves in a region of very low density and reduced dynamical friction (the so-called ``core''). 

In addition, because the two black holes are orbiting around the central massive clump, 
their relative velocity is a factor of $\sim 3$ higher compared to that in the 
standard $2$~pc simulation in which the central clump was not resolved. 
The dynamical friction force scales as $\rho/v_{\rm BH}^2$. The combined reduction
of $\rho$ and increase of $v_{\rm BH}$ results in an overall decrease of almost a factor of $50$ in the
strength of dynamical friction, explaining the observed suppression in the orbital decay by almost two orders of magnitude.
Moreover, as a result of the clump formation, the mass contained within the orbit of the two black holes 
has become much larger than the sum of their masses; the two black holes do not 
form a binary as in the $2$~pc resolution simulations but only a loose pair.

Should we trust the formation of the massive clump and the suppression in the orbital decay?
Probably not. Gas inflows are expected in non-axisymmetric disks, and they are indeed reported 
in high-resolution simulations of nuclear disks starting from equilibrium conditions (Escala 2006; 2007; Kawakatu and Wada 2008). 
Yet the magnitude of the inflows, hence the mass of the central clump and associated variation of the disk
density profile, is likely exaggerated by the crude modeling of the ISM, and even more by the lack of star
formation, supernovae feedback, and gas accretion onto the SMBHs. We take a closer look at this issue below.
In Escala (2006; 2007) and Kawakatu \& Wada (2008) the mass that collects in the inner parsec after a few 
million years is less than $1\%$ of the total mass of the nuclear disk, while it is nearly $10\%$ in our 
simulations (we note, however, that our nuclear disk is more than an order of magnitude more massive 
than the disk models used in these studies, hence stronger non-axisymmetric torques are expected due 
to the stronger self-gravity).

\section{Missing physics; multi-phase ISM, star formation, and gas accretion}

As explained in the previous sections we used an effective adiabatic equation of state with $\gamma=7/5$ to 
describe the gas thermodynamics. As discussed in M07, this equation of state breaks down
at $\rho > 10^{4}$ atoms cm$^{-3}$ because the gas becomes nearly isothermal at such high densities 
($\gamma \sim 1.1$ based on Spaans \& Silk 2000). Such high densities are indeed reached in the inner few 
parsecs of the nuclear disk. Allowing the equation of state to become softer, namely adopting a lower value of $\gamma$, 
would result in a more compressible gas at these densities; this would
in principle exacerbate the sinking problem by producing an even more concentrated profile and massive central
clump. However, the problem is that the approach of the effective equation of state becomes increasingly
less robust as the resolution is increased. With $0.1$~pc resolution we should be able to resolve very well 
the clumpy nature of the ISM and a multi-phase model would be required. 

Wada \& Norman (2001) showed that when the main radiative heating/cooling processes and supernovae feedback are 
directly incorporated in simulations of rotating gaseous disks a multi-phase, turbulent 
ISM arises naturally via a combination of gravitational instability, thermal instability, 
and turbulent energy injection by supernovae explosions. 
The resulting structure is filamentary and clumpy at all scales (Wada 2004). 
The large scale, coherent spiral patterns seen in our simulations would be replaced by much more irregular
structures with higher degree of small scale turbulence (turbulence is present in our disks, but is 
generated only at large scales by global non-axisymmetric modes). This should give rise to a stochastic, episodic
inflow (Escala 2006; Kawakatu  \& Wada 2008) as opposed to the steady inflow 
that we observe in our simulations since the coherence of large scale torques would be partially 
disrupted. 

Apart from the absence of a multi-phase model for the ISM, there are other key ingredients missing from our simulations
that would certainly have an effect on the structural evolution of the nuclear disk and thus on the sinking of 
the two SMBHs: star formation and gas accretion. Let us first consider the issue of star formation. 
During the short timescale probed by our simulations after the merger ($\sim 10^6$~yr) most
of the nuclear gas should not be converted into stars, even assuming the high 
star formation rates observed in powerful ULIRGs.
To estimate the expected star formation rate, we can simply assume that most of the mass in the nuclear disk is
molecular (as expected by the high densities, above $10^3$ atoms/cm$^3$).
This gas will be turned into stars on the local dynamical timescale. Star formation in molecular clouds
is rather inefficient; for giant and large molecular clouds ($> 10$~pc) the fraction of gas that is converted into stars 
can be as low as $1-2\%$ (Krumholz \& Tan 2007), for cloud cores, the densest regions of clouds that collapse directly 
into individual stars, it can be at most $30\%$ (Li et al. 2005) (radiative feedback and turbulence driven by supernovae 
explosions, outflows and large-scale gravitational instability all contribute to regulate the efficiency at various scales).

Let us now consider the worst case scenario which is to adopt the highest value of the efficiency 
(our spatial resolution of $2$~pc is intermediate between the scale of molecular clouds and that of their cores) and
write the star formation rate in the nuclear disk as  $dM_*/dt = 0.3 \times M_{\rm gas}/T_{\rm orb}$,
where $T_{\rm orb} = 10^6$~yr, the orbital time at the disk half mass radius of $25$~pc and 
$M_{\rm gas} = 3 \times 10^9 \Mo$.
The resulting star formation rate is $900 \Mo$/yr. Nonetheless, even with such high star formation rate less than 1/5
of the gas in the disk, $4.5 \times 10^8 \Mo$, would be converted into stars during the time 
required for the black holes to sink and bind in the nuclear disk ($5 \times 10^5$~yr). 
However, while this statement is true for the disk as a whole, if we restrict ourselves to the inner few parsecs then the
gas has such high densities that the local star formation rate could convert a few times $10^8 \Mo$ 
of gas into stars before the black holes reach a parsec scale. The rapid conversion into stars would weaken the non-axisymmetry
of the gas by reducing the gas surface density, thus increasing 
the disk Toomre $Q$ parameter and stabilizing the disk. With a weaker
spiral pattern the central gas inflow would be reduced. Kawakatu \& Wada (2008)
conclude that star formation consumes a significant fraction of the gas available in the nuclear 
disk, limiting the amount of gas that can feed the central parsec region and thus the eventual growth 
of the SMBHs. Dotti et al. (2007) showed that, once a nuclear disk
has formed, a nearly  identical decay rate of the two massive black holes follows for entirely gaseous
or stellar nuclear disks since
dynamical friction has a similar strength in gaseous and stellar backgrounds in the supersonic
regime (Escala et al. 2004). Therefore, if the central region of the nuclear disk converts rapidly into
stars this would not lower the efficiency of the decay process by itself; rather, the slowdown
of the decay might be avoided by limiting the accumulation of the dense central gas clump.

As far as gas accretion onto the black holes is concerned, part of the cold molecular gas in a 
multi-phase medium will be accreted from the surrounding nuclear disk.
This phenomenon could help with the observed black hole stalling in two ways, 
i.e. increasing the strenght of dynamical friction as the black hole become more massive 
and/or making the disk more stable by 
lowering its gas surface density. However, the ultimate effect of gas accretion can
only be elucidated with an appropriate simulation. If the two black holes manage to sink below $1$~pc and decrease 
their relative velocity they could eventually reach the ellipsoidal torque regime (Escala et al. 2004; 2005)
and continue to sink to $0.1$~pc and below.
A new ISM model which combines the equilibrium effective model of Spaans \& Silk (2000) with non-equilibrium
processes such as shock heating and supernovae explosions has been recently implemented in GASOLINE
(Roskar et al., in prep.). This model produces a multi-phase ISM with properties similar to those seen in Wada \& Norman (2002)
and Escala (2006; 2007) but it incorporates a more realistic modeling of the balance between heating
and cooling for the high density, cold gas phase based on radiative transfer calculations.
The calculations of M07 are currently being recomputed with this new model which also includes
star formation. 

\begin{acknowledgements}
We are grateful to our collaborators Simone Callegari, Monica Colpi, Piero Madau, Tom Quinn, 
Rok Roskar, and James Wadsley for allowing us to present results in advance of publication.
We acknowledge discussions with Mandeep Gill, David Merritt, and Marta Volonteri.
S. Kazantzidis is supported by the Center for Cosmology and
Astro-Particle Physics (CCAPP) at The Ohio State University.
A. Escala is funded by the U.S. Department of Energy through a KIPAC Fellowship 
at Stanford  University and the Stanford Linear Accelerator Center.
All simulations were performed on Lemieux at the Pittsburgh Supercomputing 
Center, on the Zbox and Zbox2 supercomputers at the University of Z\"urich,
and on the Gonzales cluster at ETH Z\"urich.
\end{acknowledgements}

\end{document}